\documentclass[aps,prb,citeautoscript,twocolumn,groupedaddress]{revtex4}
\usepackage{graphicx}
\usepackage{caption}

\begin{document}
\title{Collective Excitations and Thermodynamics of Disordered State: New Insights into an Old Problem}
\author{V. V. Brazhkin$^{1}$}
\author{K. Trachenko*$^{2}$}
\address{$^1$ Institute for High Pressure Physics, RAS, 142190, Moscow, Russia}
\address{$^2$ School of Physics and Astronomy, Queen Mary University of London, Mile End Road, London, E1 4NS, UK, email k.trachenko@qmul.ac.uk, tel. +440278826547}

\begin{abstract}
\begin{center}
{\bf ABSTRACT}
\end{center}
Disorder has been long considered as a formidable foe of theoretical physicists in their attempts to understand system's behavior. Here, we review recently accumulated data and propose that from the point of view of calculating thermodynamic properties, the problem of disorder may not be as severe as has been hitherto assumed. We particularly emphasize that contrary to the long-held view, collective excitations do not decay in disordered systems. We subsequently discuss recent experimental, theoretical and modelling results related to collective excitations in disordered media, and show how these results pave the way to understanding thermodynamics of disordered systems: glasses, liquids, supercritical fluids and spin glasses. An interesting insight from the recent work is the realization that most important changes of thermodynamic properties of the disordered system are governed only by its fundamental length, the interatomic separation. We discuss how the proposed theory relates to the previous approaches based on general many-body statistical mechanics framework.
\end{abstract}

\maketitle


\section{Introduction: the problem and the choice}

Heat capacity of matter is widely perceived to be one of its main important properties because it informs a scientist about the degrees of freedom available in the system. One of the first statistical physics results that a science student learns is the Dulong-Petit law: the constant-volume heat capacity of a classical harmonic solid, $C_v$, is $C_v=3Nk_{\rm B}$, where $N$ is the number of particles \cite{lanstat}. This result is explained by the possibility of representing the energy of a crystalline solid as a sum over $3N$ independent collective waves, each with the energy of $k_{\rm B}T$. The student further learns that to a very good degree, the Dulong-Petit law universally agrees with experiments on classical solids.

This textbook success story quickly fades away when we try to extend this line of thinking beyond ordered crystals, and consider two other very common examples of condensed-matter systems with {\it disorder}: amorphous glasses and liquids. We discuss the main reason for this below, and here we note the regrettable result: physics textbooks have very little, if anything, to say about their specific heat, including textbooks dedicated to liquids and other disordered systems \cite{lanstat,ziman,l1,l2,march1,l3,l4,l5}. In an amusing story about his teaching experience, Granato recalls living in fear about a potential student question about liquid heat capacity \cite{granato}. Observing that the question was never asked by a total of 10000 students, Granato proposes that ``...an important deficiency in our standard teaching method is a failure to mention sufficiently the unsolved problems in physics. Indeed, there is nothing said about liquids [heat capacity] in the standard introductory textbooks, and little or nothing in advanced texts as well. In fact, there is little general awareness even of what the basic experimental facts to be explained are.''

The main perceived reason for the notable theoretical failure to extend the derivation of the Dulong-Petit law to disordered systems such as liquids is that they are not amenable to the common harmonic approximation, the work-horse of the solid-state theory.

Two common types of disorder include static and dynamic disorder. For example, the disorder in glasses and spin glasses is static. Liquids have another important type of disorder, the dynamic disorder. It is the combination of dynamic disorder and strong interactions in liquids that has been widely viewed as an ultimate obstacle to constructing a general theory of liquids \cite{lanstat}. Indeed, interactions in a liquid are strong, and are similar to those in solids. Strong interactions are successfully treated in solids in the phonon approach, but this approach has long been thought to be inapplicable to liquids where atomic displacements are large. Stated differently, the ''small parameter'' in the theory of solids is atomic displacements, and a harmonic contribution to the energy, the phonon energy, is often a good approximation. In gases, the small parameter is weak interatomic interactions. On the other hand, liquids have none of these because displacements are large and interactions are strong at the same time.

The absence of a small parameter was, in Landau view, the fundamental property of liquids that ultimately precluded the construction of a theory of liquids at the same level existing for solids or gases. In a similarly discouraging spirit, Landau and Lifshitz Statistical Physics textbook \cite{lanstat} states twice (paragraph 66, page 187 and paragraph 74, page 222) that strong interactions, combined with system-specific form of interactions, imply that liquid energy is strongly system-dependent, precluding the calculation of energy and other properties in general form, contrary to solids or gases.

The big problem for theorists \cite{lanstat,ziman}, the disorder has been irrelevant for the experimentalists who have informed us that at high temperature, $C_v=3Nk_{\rm B}$ holds well for both glasses and liquids close to the melting point. It becomes immediately tempting to interpret this result in the same way as for crystals, i.e. on the basis of $3N$ collective modes. On the other hand, the inability to extend the method of decomposition of the system excitations into harmonic collective modes is perceived as the first general argument against such an interpretation. Next, we often read that collective excitations are strongly damped in the disordered structures of liquids and glasses \cite{egami,ruocco}. This is in the seeming agreement with numerous experimental results where the damping of waves (e.g. sound and electromagnetic waves in a wide frequency range) is widely observed. Worse still, damping is believed to increase for large wavevectors ${\bf k}$, and becomes very large at wavelengths comparable to interatomic separations, i.e. exactly in that region of $k$-space which gives the most contribution to the phonon energy (in the solid-state theory, this follows from the $\omega=kc$ relationship and associated quadratic vibrational density of states, $g(\omega)\propto k^2\propto\omega^2$). This ostensibly becomes fatal for interpreting the Dulong-Petit law in liquids on the basis of $3N$ collective modes as in crystals.

We are therefore faced with an interesting and important choice. The first approach is to search for a thermodynamic theory of disordered condensed matter systems such as liquids that explains the experimental result, $C_v=3Nk_{\rm B}$, but is essentially divorced from $3N$ collective excitations discussed in a theory of crystals. This approach should involve developing new concepts as well as continued perseverance, in view that no such theory has appeared yet at the level comparable to what we have for crystals despite over a century of intensive research.

An alternative is to state that $3N$ non-decaying (non-damped) collective modes exist in disordered systems just as they do in crystals, explore their propagation and subsequently discuss how these modes govern system's thermodynamics. It is this approach that we adopt and explore below in our quest of solving the problem of constructing a theory of disordered state. Our approach is in notable resemblance to the method of ``radical conservative-ism'' coined by Wheeler, who proposed to ``...insist on adhering to well-established physical laws (be conservative), but follow those laws into their most extreme domains (be radical), where unexpected insights into nature might be found'' \cite{radicon}.

We start with a somewhat pedagogical discussion of collective excitations to be useful to students, and subsequently proceed to discussing issues at the forefront of current condensed matter physics research, thermodynamic theory of disordered state.

\section{Damped or not damped?}

The physics of condensed matter, the system of a large number of interacting particles, includes two main tasks. The first task is to calculate the ground state: find the particle states in the phase space corresponding to the energy minimum, calculate this energy and other physical quantities. The second task is to describe the excitations in the system. Due to interactions between particles, low-energy excitations are related to collective processes. In many cases, each excitation can be described as a wave. The most commonly discussed excitation process of interacting many-body system is a harmonic plane wave describing density variations, with a certain wave vector {\bf k}. This representation describes excitations in any interacting system of atoms, electrons, spins and so on in the region of large wavelengths where structural details are insignificant and where the continuum approximation can be made.

For wavelengths comparable with interatomic separations where structural details become relevant, the continuum approximation does not apply. Consequently, the collective excitations can not be represented by harmonic waves with certain {\bf k} in general case. Nevertheless, due to the presence of periodicity in an ideal harmonic crystal, collective excitations can be represented by harmonic oscillations, plane waves, with {\bf k} corresponding to the entire possible range of the wavelengths $\lambda$, from the size of the crystal to the shortest interatomic separation. In such an ordered system, excitations are describable in terms of quasi-particles, phonons and magnons in structural or magnetic systems. Each frequency of the excitation, $\omega$, corresponds to a certain wavelength and a momentum (a quasi-momentum in a periodic system), with a relationship between $\omega({\bf k})$ and ${\bf k}$ called the dispersion relationship.

Interatomic interactions in real crystals are not entirely harmonic, and the collective excitations are no longer represented by a simple set of harmonic plane waves. To describe the effects of anharmonicity, multi-phonon processes are introduced, where, for example, one phonon decays into two, and where harmonic phonons have a finite lifetime. The reasons for finite phonon lifetimes can also include imperfections of the crystal lattice: defects, local strains and so on. Notably, decay of phonons in real crystals has important physical consequences including, for example, finite thermal conductivity. Decay and lifetimes of phonons, magnons and other collective excitations is routinely measured experimentally using scattering of neutrons, photons and other elementary particles.

The absence of periodicity in disordered media (glasses, liquids, spin glasses and so on) precludes the representation of collective excitations in terms of harmonic plane waves with a set of ${\bf k}$. If we continue to use plane waves as an approximation, we encounter decay and finite lifetimes. These turn out to be small for large wavelengths where the the continuum approximation applies. For smaller wavelengths, decay of phonons increases, and is commonly seen as widening of the intensity of the dynamic structure factor in either $\omega$ or $k$. For smallest wavelengths comparable to interatomic separations, the effective phonon lifetime can become smaller than the inverse frequency. For this reason, we might wonder to what extent we can talk about high-frequency excitations in liquids and glasses. Indeed, textbooks and recent research papers state that the concept of phonons or phonon-like modes in liquids is either questionable \cite{egami} or not well-defined due to strong damping \cite{ruocco}. Our analysis below shows that this is not the case. Our main point is that collective excitations that are {\it eigenstates}, do not decay in any system, ordered or disordered.

Quite generally, lets consider a system of particles where a state of thermal equilibrium has been previously achieved (due to contact with a thermal bath and non-linear effects leading to energy distribution between modes) so that the system temperature can be introduced. Let us now isolate this system from its environment and consider the system at given thermodynamic parameters such as temperature and pressure. The energy of such a closed system is conserved. We do not consider fine and small-magnitude effects such as quantum interactions of system particles with zero oscillations of the vacuum and so on. Because the energy is conserved, there are no channels for energy dissipation. Consequently, collective excitations in condensed matter systems do {\it not} decay. This is illustrated in the following, more specific, example.

Let us consider a classical topologically disordered, amorphous, system of $N$ classical particles, the ``balls and springs'' model, and assume that interactions are well-described by the harmonic quadratic interaction. This system can be solved exactly \cite{landau}, with the result that the system has $3N$ normal modes with frequencies ${\omega_i}$, linear superposition of which give the collective excitations in the disordered system. Being linear combinations of eigenstates, the collective excitations obviously do {\it not} decay (here and below, we will refer to these excitations as eigenstate collective excitations, ECEs). Note that at no stage the solution relies on the existence of ordered (or disordered, for that matter), system. In this model, the periodicity is introduced as the next step only in order to calculate a dispersion relationship for a particular ordered system using Fourier transforms.

The only difference between this model and the ordered system is the absence of a well-defined relationship between $\omega$ and ${\bf k}$ inasmuch as ${\bf k}$ is not defined well due to the lack of periodicity, especially at small wavelengths. Nevertheless, this difference bears no significance for the question whether collective excitations decay or not. Collective excitations in our system, being the eigenstates, do not decay. Moreover, because there is a well-defined set of eigenmode frequencies $\omega_i$, we can rigorously introduce the vibrational density of states $g(\omega)$. This fact will come important below when we discuss calculating thermodynamic functions of disordered systems.

Lets expand the conditions imposed on our disordered system, and consider a complex interaction that is not quadratic. In this case, the ECEs can not be represented by simple harmonic functions, and may acquire a complicated mathematical description. The resulting collective excitations can no longer be represented by a superposition of non-interacting normal modes. Yet even in this case, these ECEs do not decay in our conservative system.

We therefore conclude that contrary to what is often stated \cite{egami,ruocco}, collective excitations in disordered condensed matter are well-defined and are not damped. These excitations may not be described by plane waves with fixed ${\bf k}$ as in ordered systems, yet importantly, this circumstance is largely irrelevant for constructing a thermodynamic theory of disordered systems as discussed below.

\section{Living with disorder}

Lets now externally introduce a set of plane harmonic waves with well-defined energy and ${\bf k}$ in our disordered system balls and springs system with quadratic interactions and follows its propagation. This set will transform into excitations with different frequencies ${\omega_i}$ traveling in different directions. This gives decay of plane waves, and sets the ``decay time'', time during which ECEs can be represented by plane harmonic waves. In disordered systems at large wavelengths where the continuum approximation works well, this time can be quite long and, consequently, decay and damping are small. At wavelengths smaller than the medium-range order where the details of structure become important (about 10--20 interatomic distances), this time is small and damping is large.

Experimentally, the measured decay of phonons and other quasi-particles is related to the fact that the external probes (photons, electrons, neutrons) are harmonic plane waves with well-defined $\omega$ and ${\bf k}$. These external probes interact with the ECEs in disordered media, and register the time during which these excitations resemble a plane harmonic excitation. It is this time that is often taken as the measure of decay of the plane harmonic excitation in disordered systems.

From this perspective, we encounter three different cases:

\begin{enumerate}
\item Plane harmonic waves in ordered systems (ECEs): Non-decayed and Not Damped
\item Plane harmonic waves in disordered systems (not ECEs): Decayed and Damped
\item ECEs (non-plane waves) in disordered systems: Non-decayed and Not Damped
\end{enumerate}

Historically, the approaches to disordered systems were mainly based on case (2) above, and involved new concepts and ideas emerging in condensed matter physics. For example, the so-called Ioffe-Regel limit is related to the transition from propagating to non-propagating phonons in disordered systems, and has been widely discussed to study decay and attenuation in glasses. Using approach (2), several classification methods have been proposed that are based on different decay of harmonic plane waves in disordered media at large and small wavelengths. For long wavelengths and small energies, ECEs in a disordered system are similar to plane-wave harmonic phonons, whereas at small wavelengths, they differ from the plane waves as discussed above. Although rather loose, this distinction into ``good'' and ``bad'' phonons has formed the common picture of collective excitations in disordered systems. This picture has attracted a considerable amount of research. For example, ``extendons'' and ``diffusons'' have emerged as new terms to refer to ``bad'' phonons in disordered media \cite{extend,diffus}.

The dissimilarity of collective excitations in disordered systems from plane waves bears notable implications for several physical phenomena. For example, transport effects such as thermal conductivity is reduced in disordered systems as compared to crystals, and its temperature dependence differs in two systems. At the same time - and this is the central point of this paper - thermodynamic properties of disordered systems turn out to be insensitive to disorder and to the difference of the ECEs in disordered systems from the harmonic plane waves matters. Contrary to the long-held belief \cite{ziman}, recent experimental and theoretical work suggests that disorder is not an issue from the point of view of thermodynamics.

Fairly recently accumulated experimental data, primarily from inelastic X-ray and neutron scattering in disordered systems, reveal an important result: in amorphous solids, glasses, $\omega$ continues to be approximately proportional to $k$ as in the crystal with the effective speed of sound being close to the speed of sound of long-wavelength phonons. The same is the case in liquids, where recent experiments informed us that a well-defined dispersion relationship exists in disordered systems up to the largest $k$ corresponding to the shortest interatomic separation in the system \cite{burkel,pilgrim,mon-ga,mon-na,hoso,water,disu1,disu2,ruzi,baldi}.

Before we review these experiments, we note an interesting and important difference between collective modes in glasses and liquids, first proposed by J. Frenkel \cite{frenkel}. Frenkel introduced liquid relaxation time $\tau$ as the average time between particle jumps at one point in space in a liquid, and subsequently identified $\tau$ with the phenomenological time constant in the Maxwell relationship $\eta=G_{\infty}\tau$, where $\eta$ is liquid viscosity and $G_{\infty}$ is the instantaneous shear modulus (which can be taken as the shear modulus at high finite \cite{puosi} frequency). Each particle jump corresponds to the transition between two quasi-equilibrium liquid configurations. Stillinger and Weber observed these transitions in molecular dynamics simulations a few decades ago \cite{stil}. They found that each transition separates local energy minima forming the multitude of ``inherent structures'' in the liquid.

The range of $\tau$ is bound by two important values: at low temperature, $\tau$ increases until it reaches the value at which the liquid stops flowing at the experimental time scale, corresponding to $\tau\approx 10^{3}$ s and the liquid-glass transition \cite{dyre}. At high temperature, $\tau$ approaches its minimal value given by Debye vibration period, $\tau_{\rm D}\approx 0.1$ ps, when the time between the jumps becomes comparable to the shortest vibrational period.

With a remarkable physical insight, Frenkel developed the argument about the liquid vibrational states as follows. At times shorter than $\tau$, a liquid is a solid, and therefore supports one longitudinal mode and two transverse modes, whereas at times longer than $\tau$, liquid flows and loses its ability to support shear stress, and therefore supports one longitudinal mode only as any elastic medium (in a dense liquid, the wavelength of this mode extends to the shortest wavelength comparable to interatomic separations as discussed below). This is equivalent to asserting that the only difference between a glass and a liquid is that a liquid does not support all transverse modes as the solid glass does, but only those with high frequency $\omega>\frac{1}{\tau}$. The longitudinal mode is unaffected in this picture except for different dissipation laws in the regimes $\omega>\frac{1}{\tau}$ and $\omega<\frac{1}{\tau}$ \cite{frenkel}.

Derived on purely theoretical grounds, Frenkel's idea was later experimentally confirmed. Liquids and, notably, low-viscous liquids such as Na and Ga, have been shown to support transverse modes with frequency reaching the highest Debye frequency \cite{mon-ga,mon-na}. This is an important assertion because high-frequency modes give the largest contribution to energy and heat capacity. Interestingly, the experimental evidence for high-frequency transverse modes in liquids has arrived more than 70 years after Frenkel's prediction \cite{mon-ga,mon-na}. Such a time lag could be thought of as surprisingly long, however mapping dispersion curves for these and other liquids required powerful synchrotron radiation sources which started to be deployed relatively recently. The long-lived absence of experimental evidence for high-frequency transverse modes in liquids has probably contributed to the lack of their theoretical understanding discussed in the Introduction.

Since fairly recently, the ability of liquids to support solid-like collective modes with wavelengths extending to the shortest distance comparable to interatomic separations has been widely ascertained experimentally on the basis of measured dispersion curves \cite{burkel,pilgrim}. This includes transverse modes in both highly-viscous \cite{w3} and low-viscous liquids such as Na and Ga, water and so on \cite{mon-ga,mon-na,hoso,water}. Solid-like dispersion relationships have also been found in supercritical fluids well above the critical point \cite{disu1,disu2}.

As an illustrative example, we show recently measured dispersion curves for liquid Ga \cite{mon-ga} and liquid Na \cite{mon-na}, together with SiO$_2$ glass \cite{ruzi,baldi} for comparison in Figure \ref{1}. We observe a striking similarity in dispersion curves between the liquids and their polycrystalline and/or bulk counterparts including both longitudinal and transverse modes. Together with dispersion curves seen in SiO$_2$ glass, Figure \ref{1} presents an important experimental evidence regarding collective excitations: despite topological disorder in glasses and combined topological and dynamic disorder in liquids, solid-like quasi-linear dispersion curves exist in these systems in a wide range of $k$ and up to the largest $k$ corresponding to interatomic separations, as in crystals.

\begin{figure}
\begin{center}
{\scalebox{0.4}{\includegraphics{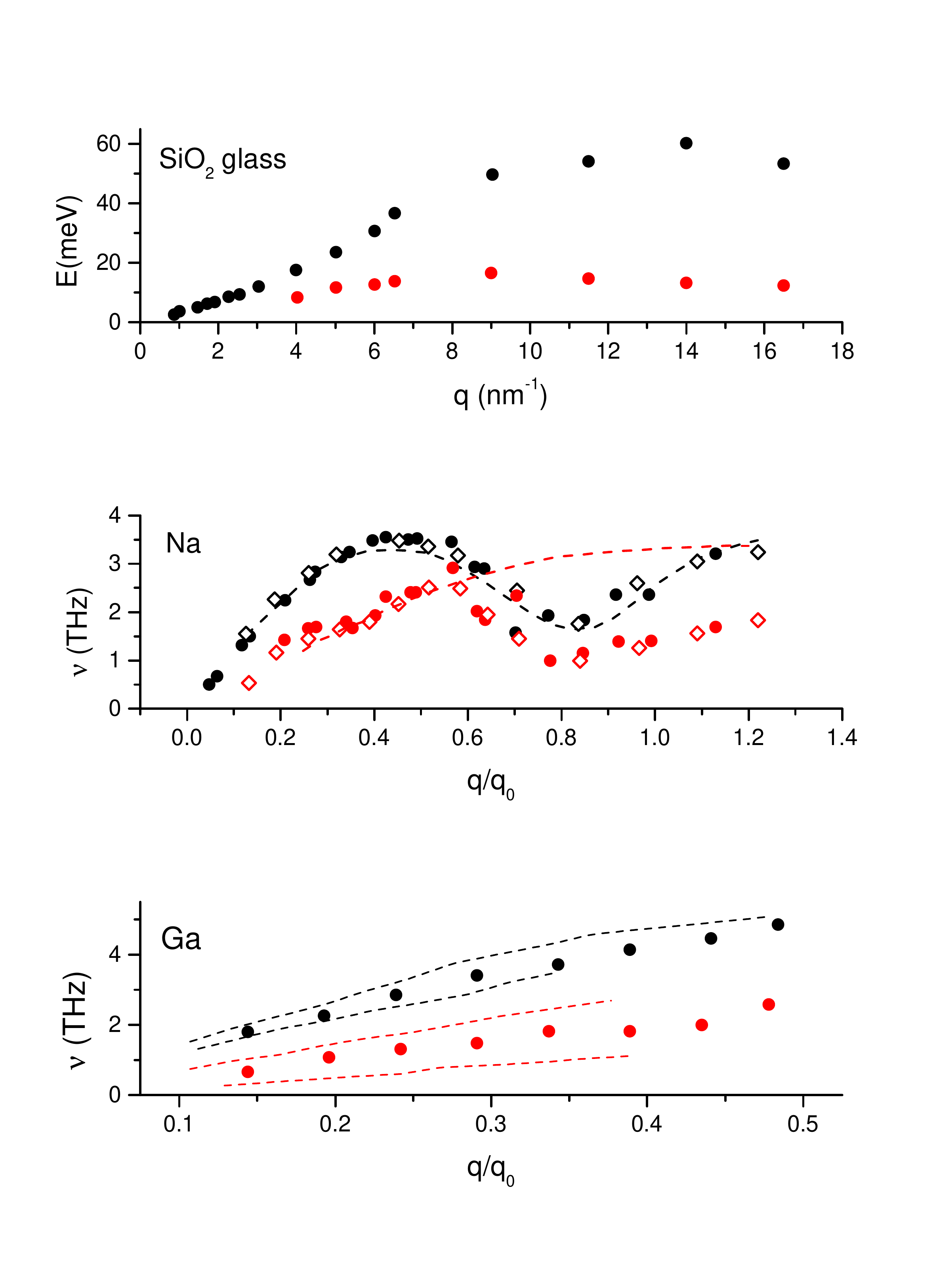}}}
\end{center}
\caption{Top: longitudinal (filled black bullets) and transverse (filled red bullets) dispersion curves in SiO$_2$ glass \protect\cite{ruzi}. Middle: longitudinal (filled black bullets) and transverse (filled red bullets) excitations in liquid Na. Open diamonds correspond to longitudinal (black) and transverse (red) excitations in polycrystalline Na, and dashed lines to longitudinal (black) and transverse (red) branches along [111] direction in Na single crystal \protect\cite{mon-na}. Bottom: longitudinal (black bullets) and transverse (red bullets) excitation in liquid Ga. The dots are bracketed by the highest and lowest frequency branches measured in bulk crystalline $\beta$-Ga along high symmetry directions, with black and red dashed lines corresponding to longitudinal and transverse excitations, respectively \protect\cite{mon-ga}. Dispersion curves in Na and Ga are reported in reduced zone units.}
\label{1}
\end{figure}

Physically, the existence of dispersion curves in disordered systems as probed by the harmonic probes such as X-rays or neutrons is due to the fact that despite disorder, a well-defined short-range order exists in these systems as is seen in the peaks of pair distribution functions in the short (as well as medium) range. Therefore, high-frequency harmonic plane waves, even though damped, are able to propagate at least the distance comparable to the typical length of the short-range order. We will find below that this length, the interatomic separation (which is also the fundamental length of the system), plays a profound role in governing thermodynamic properties of disordered systems.

Notably, most of the experimental evidence of propagating modes in liquids, such as that illustrated in Figure \ref{1}, is fairly recent, and has started to come to the fore only when powerful synchrotron radiation sources started to be deployed, some 50--60 years after Frenkel's prediction. This long-lived absence of experimental data about propagating collective excitations in liquids may have contributed to their poor understanding from the theoretical point of view.

The striking similarity of dynamics in liquids and their crystalline counterparts discussed above interestingly compares to the recently discovered similarity of dynamics between glasses and crystals in terms of the ``Boson'' peak, the widely discussed peak in the reduced density of states at low frequency \cite{chumakov}. The Boson peak has been long thought to be present in glasses only but not in crystals and originate as a result of disorder, posing an intriguing general question of how disorder affects dynamics. Recent work \cite{chumakov} has demonstrated that similar vibrational features are present in crystals as well, provided glasses and crystals have similar density.

\section{Thermodynamics of disordered systems: glasses, liquids, supercritical fluids and spin glasses}

\subsection{Calculating the energy of a disordered system}

We have ascertained above that a disordered system is capable supporting non-decaying, propagating, collective excitations, just as a crystal is. The experiments have made a further assertion: these excitations can be well represented by plane waves that extend to the shortest wavelength comparable to interatomic separation. We now write the vibrational energy of the disordered system as

\begin{equation}
E=\int E(\omega,T)g(\omega)d\omega
\label{1}
\end{equation}

\noindent where $E(\omega,T)=T$ or $E(\omega,T)=\frac{\hbar\omega}{2}+\frac{\hbar\omega}{\exp\frac{\hbar\omega}{T}-1}$ is the energy of classical and quantum harmonic oscillator. Here and below, $k_{\rm B}=1$.

Eq. (1) is general, and makes no explicit reference to order or disorder. The order, or the lack of thereof, is contained in $g(\omega)$ only. Looking at Eq. (1), we recognize that in order to calculate the energy using, for example, the Debye model, we {\it only} require that $\omega=ck$ approximately holds in the entire range of $k$, from the inverse of system size to the inverse of the shortest interatomic separation, the result ascertained above experimentally. The fact that plane harmonic waves differ from the ECEs of the disordered system (especially so at small wavelengths) becomes irrelevant as far as Eq. (1) is concerned. Indeed, Eq. (1) makes no reference to phonon decay process such as decay time or propagation length, and gives the same result regardless of whether decay time is measured in seconds or picoseconds as in low-viscous liquids.

A more general comment is appropriate in relation to Eq. (1): we observe that many system details, including disorder itself, are irrelevant in this calculation, and are contained in one function only, $g(\omega)$. This approach is rather typical in statistical physics where an enormously large number of degrees of freedom is substituted by a small number of thermodynamic quantities and functions \cite{lanstat}.

With regard to Debye model itself, we note that this approximation applies best not to crystals where $g(\omega)$ has many order-related features, but to the isotropic disordered systems \cite{lanstat} where these features disappear and smoothen $g(\omega)$ as a result. In this sense, approximating the dispersion relationship for disordered systems, even though scattered at large $k$ (see Figure \ref{1}), by $\omega=ck$ and obtaining the Debye density of states $g(\omega)\propto\omega^2$ may appear no worse than making the same approximation for the anisotropic crystal with potentially rich set of optic modes and features in $g(\omega)$.

We now briefly review how Eq. (1) can be applied to calculating thermodynamic properties of three most common disordered systems: glasses, liquids and spin glasses.

\subsection{Glasses}

Calculating $E$ using the Debye density of states and gives the common solid-like expression for the glass specific heat, $c_v=\frac{1}{N}\frac{dE}{dT}$, with $c_v\approx 3$ at high temperature as is widely seen experimentally (here and below, $k_{\rm B}=1)$. Slight deviations of $c_v$ from $3$ are due to intrinsic anharmonicity, as in crystals. At low temperature, $c_v\propto T^3$ as the Debye model predicts. Here, we do not discuss the contribution of the linear specific heat anomaly of glasses that operates at very low temperature of about 1 K only.

The Dulong-Petit result for glasses is perhaps not surprising, in view of widely perceived applicability of this result to all solids including amorphous ones. At the same time, this result gives experimental support to our earlier theoretical discussion about propagation and decay of collective modes in the amorphous structure. Below we review results that are much less expected and were not appreciated until very recently, results related to how collective modes can be used to calculate thermodynamic properties of liquids in a wide range of viscosity as well as supercritical fluids.

\subsection{Liquids and supercritical fluids}

There are three physically distinct regimes of liquid specific heat where $c_v$ stays close to 3 as in the solid, then decreases to 2 on temperature increases, and eventually tends to its ideal-gas value of $\frac{3}{2}$.\\

1. Viscous regime, $\tau\gg\tau_{\rm D}$.

This condition corresponds to particles oscillating many times around their quasi-equilibrium positions before jumping to the neighboring sites. It can be rigorously shown that the average liquid energy and $c_v$ are given, to a very good approximation, by the vibrational contribution only and, moreover, that this result is consistent with liquid entropy exceeding solid entropy \cite{duality}. Physically, this follows from a simple statistical-physics argument that because the relative number of diffusing atoms is $\frac{\tau_{\rm D}}{\tau}$, the energy of diffusive motion is negligible in the regime $\tau\gg\tau_{\rm D}$ \cite{duality}. This leaves vibrations as the only contributing term in the system's energy. Further, the vibrational energy of a viscous liquid in the regime $\tau\gg\tau_{\rm D}$ is, to a very good approximation, equal to the vibrational energy of the solid \cite{duality} (we note that the liquid vibrational energy may include the contribution from ``beta'' relaxation process, tentatively associated with the oscillations of cages around particles). Then, $c_v$ is close to 3 as in glasses or crystals:

\begin{equation}
c_v=3
\label{cv3}
\end{equation}

\noindent where for simplicity we have neglected the anharmonicity which generally increases $c_v$ as approximately $c_v=3(1+\alpha T)$, where $\alpha$ is the coefficient of thermal expansion \cite{glass,andri}.

Notably, in glass-forming liquids $\tau$ increases on lowering the temperature from its high-temperature limiting value of $\tau_{\rm D}\approx 0.1$ ps to $\tau\approx 10^{3}$ s at which point a liquid forms a glass, or over 16 orders of magnitude of $\tau$. Therefore, $\tau\gg\tau_{\rm D}$ and Eq. (\ref{cv3}) hold in almost entire range of $\tau$ in which liquids exist as such.

It is interesting to consider how heat capacity changes at the liquid-glass transition. At the glass transition temperature $T_g$ (temperature at which the viscous flow stops at the experimental time scale), constant-pressure heat capacity $c_p$ undergoes a jump, the main experimental signature of the glass transition. The origin of the jump and the nature of liquid-glass transition have been intensely debated for a long time with no agreement emerging \cite{dyre}. A large part of this debate concerns a question whether the liquid-glass transition and similar ``freezing'' transitions in disordered systems are related to a new class of phase transitions and critical phenomena with novel and unusual properties \cite{dyre,glass}. Given that Eq. (\ref{cv3}) predicts a constant $c_v$, it is interesting to ask how the jump of heat capacity at $T_g$ can be understood. We note that Eq. (\ref{cv3}) is the harmonic result, and that the common effect of anharmonicity is to soften phonon frequencies which, in turn, affects $c_v$ and $c_p$. An important insight comes from the observation \cite{glass} that anharmonicity affects phonon frequencies, $c_v$ and $c_p$ differently above and below $T_g$. Indeed, pressure- or temperature-induced deformation above $T_g$ includes both elastic and viscous component but has only elastic component below $T_g$ because the viscous component disappears at $T_g$ by definition (here we assume that deformation is small and is below the threshold of inducing local relaxation events \cite{window} in the solid). This gives different $\alpha$ and bulk modulus $B$ below and above $T_g$. Then, the jump of $c_v$ and $c_p$ at $T_g$ follow because the anharmonic $c_v$ can be approximately written as $c_v=3(1+\alpha T)$ \cite{glass,andri}, and $c_p=c_v+nT\alpha^2 B$ ($n$ is number density), in quantitative agreement with the experimental jump of $c_p$ \cite{glass}.\\

2. Low-viscous ``rigid'' liquid, $\tau\gtrsim\tau_{\rm D}$.

As discussed above, a liquid supports rigidity and solid-like shear modes with frequency $\omega>\frac{1}{\tau}$. When $\tau\gg\tau_{\rm D}$ as in regime (1), the system supports rigidity and most of its shear modes. As $\tau$ decreases with temperature, the number of shear modes in the system decreases; however as long as long as $\tau>\tau_{\rm D}$, rigidity exists at frequencies larger than $\frac{1}{\tau}$. We call this state of the liquid ``rigid'' liquid, to differentiate it from the ``non-rigid'' fluid state where $\tau$ approaches its minimal value of $\tau_{\rm D}$ at high temperature so that no rigid modes can be sustained at any frequency. The non-rigid gas-like fluid state is discussed in the next subsection.

In the regime $\tau>\tau_{\rm D}$, temperature increase results in the progressive decrease of the number of shear modes with frequency $\omega>\frac{1}{\tau}$ and associated decrease of the vibrational term in the total energy. To account for this effect, we can use the Debye density of states discussed above and Eq. (1) to calculate the energy of two transverse modes with frequency $\omega>\frac{1}{\tau}$ as well as the energy one longitudinal mode that is unaffected by atomic jumps in the Frenkel picture as discussed above. Adding these terms, together with the energy of diffusing atoms gives the liquid energy, $E_l$ \cite{prb1}:

\begin{equation}
E_l=NT\left(3-\left(\frac{\omega_{\rm F}}{\omega_{\rm D}}\right)^3\right)
\label{2}
\end{equation}

\noindent where $\omega_{\rm F}=\frac{1}{\tau}$ is Frenkel frequency and $\omega_{\rm D}$ is Debye frequency.

Anharmonicity and thermal expansion, particularly large in liquids, result in the extra term, $\left(1+\frac{\alpha T}{2}\right)$ \cite{glass,andri,bolm-prb}, in Eq. (\ref{2}).

According to Eq. (\ref{2}), liquid $c_v=3$ when $\tau\gg\tau_{\rm D}$, consistent with the result in regime (1) above. $c_v$ starts to noticeably decrease from 3 when $\tau$ starts approaching $\tau_{\rm D}$. When $\tau=\tau_{\rm D}$, Eq. (\ref{2}) gives $c_v=2$, corresponding to the complete loss of two shear modes and their potential energy becoming zero. This value of $c_v$, $c_v=2$ has a special physical meaning as discussed in the next subsection below.

More recently, Eq. (\ref{2}) has been extended to the quantum regime, with added effects of intrinsic anharmonicity. Taking $\omega_{\rm F}$ from independent viscosity measurements, we have found that the calculated $c_v$ agrees reassuringly well with experimental data for 21 noble, metallic, molecular and network liquids in a wide range of temperature and pressure using no free fitting parameters \cite{scirep}. Similar to predictions of Eq. (\ref{2}), $c_v$ for these liquids has been found to decrease from approximately the solid-state value of $3$ at low temperature to $2$ at high, corresponding to progressive loss of transverse modes with $\omega<\omega_{\rm F}$ with temperature increase \cite{scirep}. This universality highlights the generality in which collective modes govern thermodynamic properties of liquids. In Figure \ref{cv3} we show experimental and calculated $c_v$ for systems representing three different classes of metallic, molecular and noble liquids.\\

\begin{figure}
\begin{center}
{\scalebox{0.35}{\includegraphics{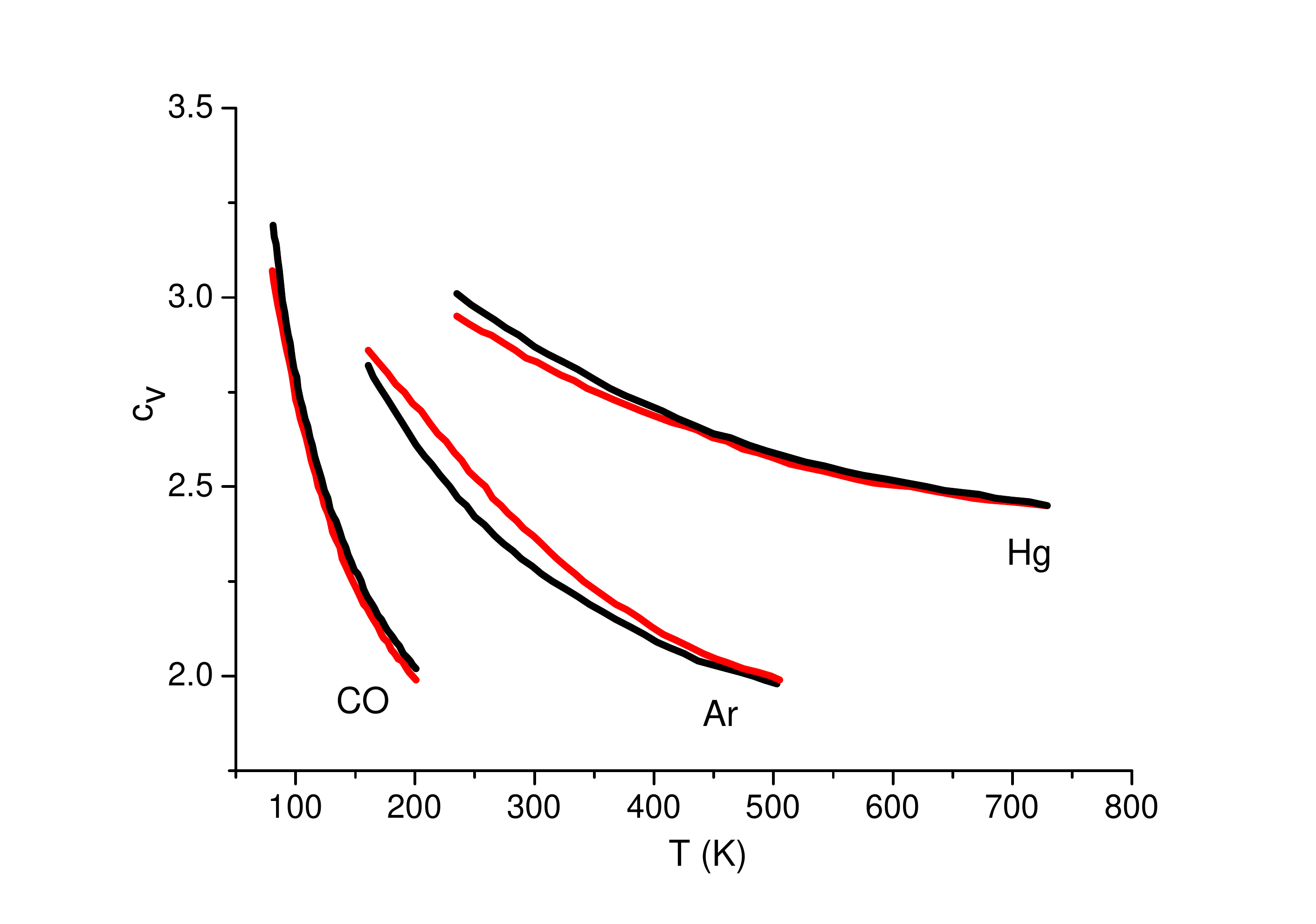}}}
\end{center}
\caption{Experimental (black color) and theoretical (red colour) $c_v$ ($k_{\rm B}=1$) for liquid Hg, Ar and CO \cite{scirep}. Theoretical $c_v$ was calculated using the quantum-mechanical analogue of Eq. (\ref{2}) with the account of anharmonicity and thermal expansion. $\omega_{\rm F}$ was calculated as $\omega_{\rm F}=\frac{1}{\tau}=\frac{G_\infty}{\eta}$, where $G_\infty$ is the infinite shear modulus and viscosity $\eta$ was taken from independent experimental data.}
\label{cv3}
\end{figure}

3. Low-viscous ``non-rigid'' gas-like fluid.

It is interesting to ask what happens when $c_v$ reaches 2 and the temperature is increased even further. On general grounds, $c_v$ should turn to its gas-like value $c_v=\frac{3}{2}$. To explore this further, we are prompted to look at conditions above the critical point where $c_v$ is not affected by the first-order boiling transition and is continuous. Accordingly, we enter the exciting realm of supercritical state, the area that has recently seen an explosion of ways in which supercritical fluids are employed in cleaning, extracting and dissolving applications yet has remained {\it terra incognita} from theoretical perspective, the unfortunate result that is also seen as limiting further industrial deployment \cite{super}.

\begin{figure}
\begin{center}
{\scalebox{0.35}{\includegraphics{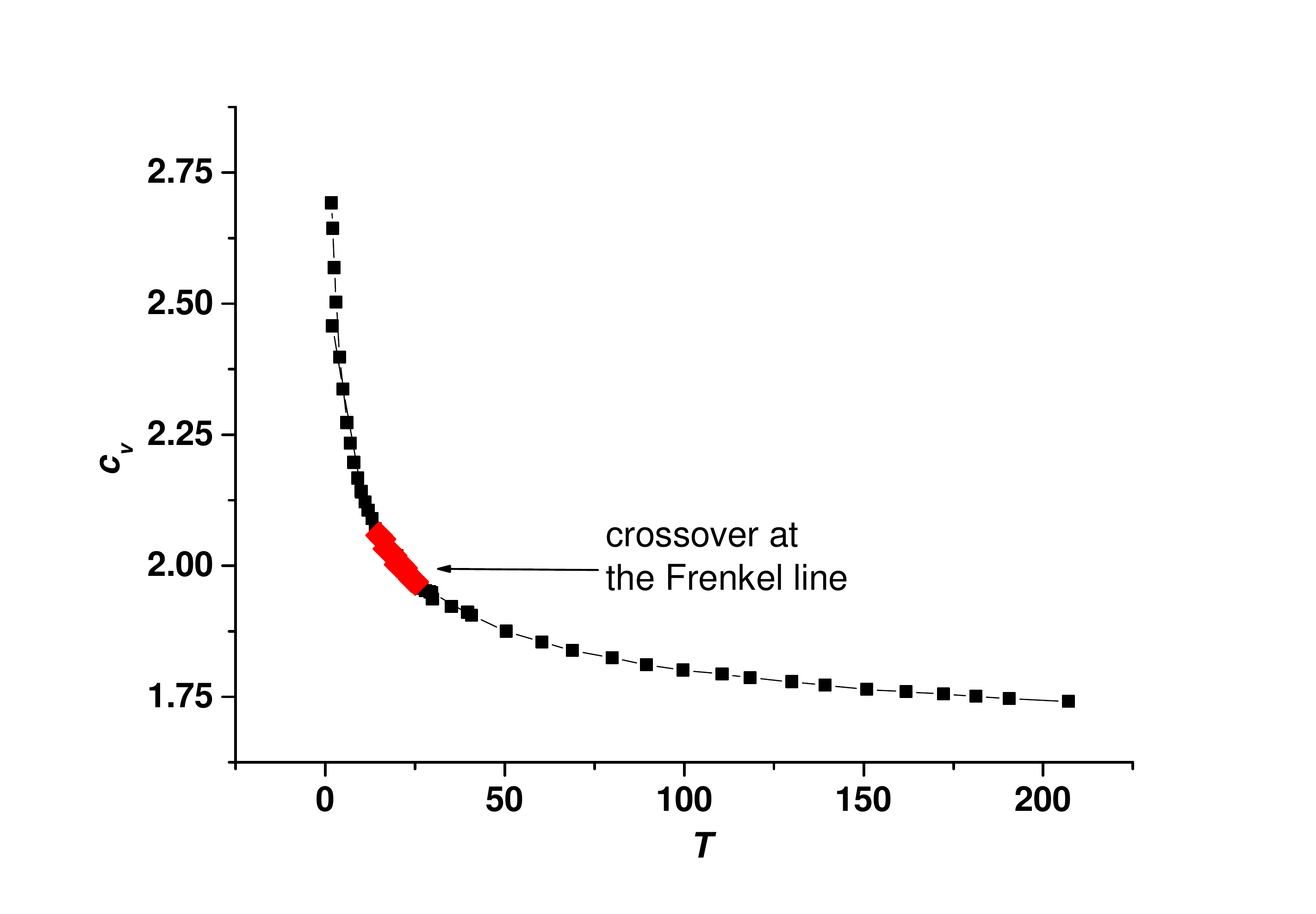}}}
\end{center}
\caption{$c_v$ ($k_{\rm B}=1$) as a function of temperature from the molecular dynamics simulation of the Lennard-Jones (LJ) liquid \protect\cite{pre}. Temperature is in LJ units. Density is $\rho=1$ in LJ units. The dynamic crossover at $c_v=2$ is highlighted.}
\label{fig2}
\end{figure}

Well above the critical point, we have found an intriguing result: $c_v$ undergoes a crossover at $c_v=2$ \cite{heat3}. We note that experimental isochoric data \cite{nist} does not extend to temperatures high enough to see the crossover, and for this reason we show $c_v$ from the molecular dynamics (MD) simulation where increasing temperature range is straightforward. In Figure \ref{fig2}, we show $c_v$ calculated using the data for the Lennard-Jones system \cite{pre}. Importantly, we observe the crossover at around $c_v=2$, the assertion supported by the data analysis in the double-logarithmic plot as in the previous report \cite{heat3}.

The crossover at $c_v=2$ is not coincidental, and takes place at the recently proposed Frenkel line which separates two distinct dynamic regimes of particle motion \cite{phystoday,pre,prl}: below the Frenkel line the motion is the combination of oscillatory and diffusive jumps, whereas above the line the oscillatory motion is lost completely, leaving the gas-like diffusive motion only. Crossing the Frenkel line above the critical point results in the qualitative change of all major properties of the supercritical fluid, including its speed of sound, viscosity, diffusion, thermal conductivity, disappearance of ``fast sound'' above the line and so on \cite{phystoday,pre}.

Important for this discussion, the Frenkel line approximately corresponds to $\tau=\tau_{\rm D}$, or to $\omega_{\rm F}=\omega_{\rm D}$, and consequently to the complete loss of the ability to support rigidity and solid-like shear modes at all available frequencies \cite{phystoday,pre}. The system therefore becomes a ``non-rigid'' gas-like fluid.

According to Eq. (\ref{2}), $\omega_{\rm F}=\omega_{\rm D}$ gives $c_v=2$, corresponding to the potential energy of two shear modes becoming zero. Importantly, the crossover of $c_v$ in Figure \ref{fig2} takes place around $c_v=2$, the point to which we return below.

As temperature is increased above the Frenkel line and $c_v$ decreases further, a new mechanism governing $c_v$ comes into play: the remaining longitudinal mode is modified in terms of its spectrum, the process that can be discussed as follows. The mean free path of the particles, $l$, increases with temperature, and defines the minimal length of the longitudinal mode because the system can not oscillate at wavelengths smaller than $l$ \cite{heat3}. The system energy can now be evaluated using experimental results \cite{disu1,disu2} that the relationship $\omega\propto k$ approximately holds even in the supercritical fluids, as is does in subcritical liquids. This implies that the Debye model developed originally for solids can be used not only for the subcritical liquids as discussed above, but also for the supercritical fluids. Consequently, the quadratic Debye density of states can be used to compute the system energy using Eq. (1). Calculating the energy of the longitudinal mode with the wavelength larger than $l$ and adding relevant kinetic terms of the diffusive motion gives the energy of the fluid, $E_f$ \cite{heat3}:

\begin{equation}\label{energy}
E_f=\frac{3}{2}NT+\frac{1}{2}NT\left(\frac{a}{l}\right)^3
\label{4}
\end{equation}

\noindent where $a$ is interatomic separation on the order of \AA, and where for simplicity we omitted the anharmonic effects.

As $l$ increases with temperature, Eq. (\ref{4}) predicts the decrease of $c_v$. Taking $l$ from independent viscosity measurements in the gas-like regime where $\eta=\frac{1}{3}\rho ul$ (here, $\rho$ is density and $u$ is the average particle speed), we have found that Eq. (\ref{4}) gives good agreement with experimental $c_v$ of several monatomic and molecular supercritical fluids in a wide temperature range with no free fitting parameters \cite{heat3}. When $l\gg a$, Eq. (\ref{4}) predicts that $c_v$ tends to its ideal-gas value of $\frac{3}{2}$ as one expects in the regime where kinetic energy dominates.

When $l\approx a$, Eq. (\ref{4}) gives $c_v=2$, matching the result from Eq. (\ref{2}) at $\omega_{\rm F}=\omega_{\rm D}$. The crossover at $c_v=2$ seen in Figure \ref{fig2} can therefore be understood to be due to two different mechanisms: at low temperature below the Frenkel line, the decrease of $c_v$ is due to the progressive loss of two transverse modes with frequency $\omega>\frac{1}{\tau}$ whereas at high temperature above the line, the decrease of $c_v$ is due to the progressive loss of remaining longitudinal mode with wavelength smaller than $l$.

\subsection{Thermodynamics of disordered systems and fundamental length}

Interestingly, the behavior of liquid $c_v$ in its entire range from the solid value, $c_v=3$, to the ideal-gas value, $c_v=\frac{3}{2}$, can be unified and generalized in terms of wavelengths.

\begin{figure}
\begin{center}
{\scalebox{0.35}{\includegraphics{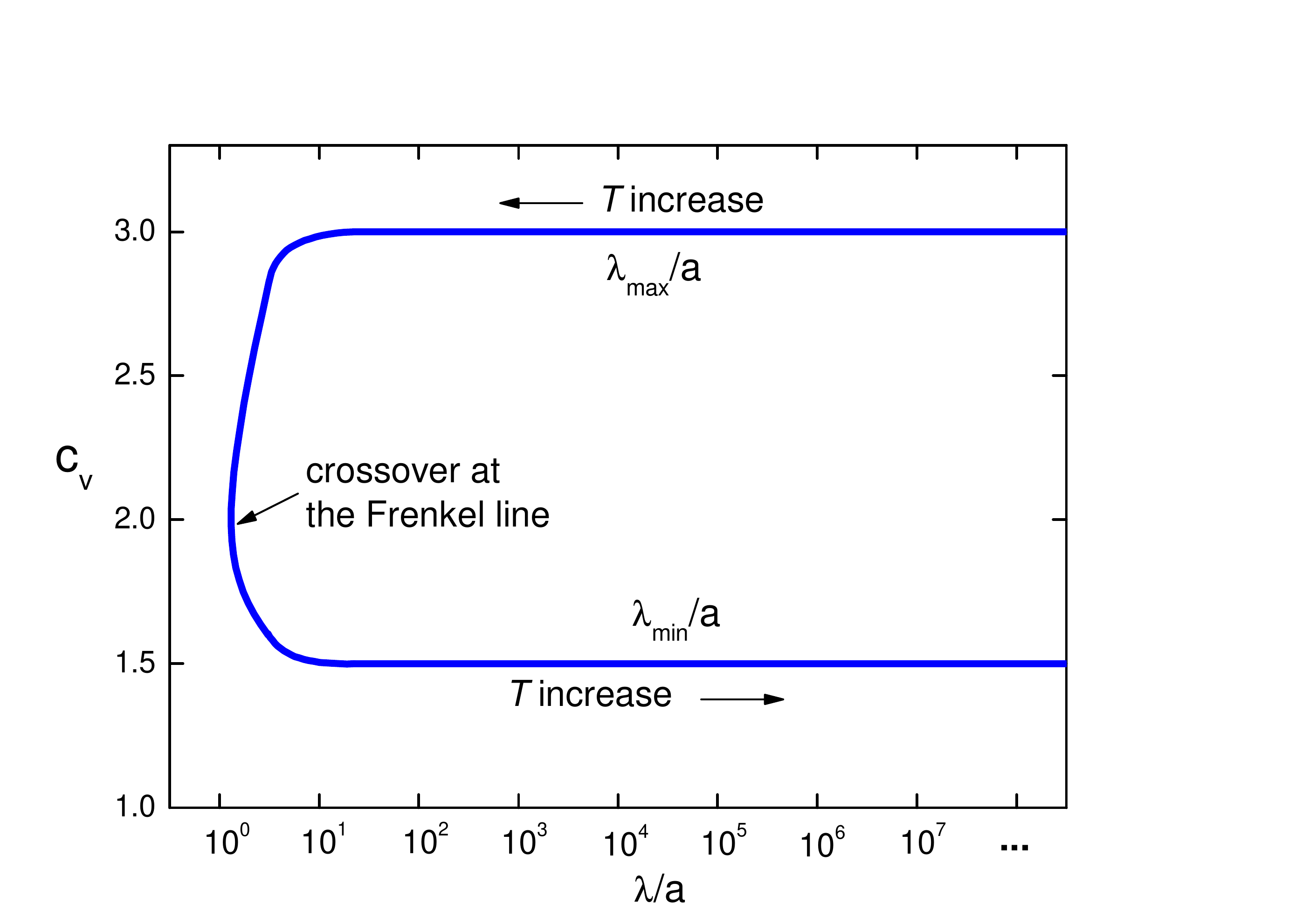}}}
\end{center}
\caption{$c_v$ as a function of the characteristic wavelengths $\lambda_{\rm max}$ (maximal transverse wavelength in the system) and $\lambda_{\rm min}$ (minimal longitudinal wavelength in the system) illustrating that most important changes of thermodynamics of the disordered system take place when both wavelengths become comparable to the fundamental length $a$.}
\label{length}
\end{figure}

An instructive insight comes from the combination of Eqs. (\ref{cv3})--(\ref{4}) once Eq. (\ref{2}) and (\ref{4}) are both interpreted in terms of wavelengths (see Figure \ref{length}). The minimal frequency of transverse modes that a liquid supports, $\omega_{\rm F}$, corresponds to the maximal transverse wavelength, $\lambda_{\rm max}$, $\lambda_{\rm max}=a\frac{\omega_{\rm D}}{\omega_{\rm F}}=a\frac{\tau}{\tau_{\rm D}}$, where $a$ is the interatomic separation, $a\approx 1-2$ \AA. According to Eq. (\ref{2}), $c_v$ remains close to its solid-state value of 3 in almost entire range of available wavelengths of transverse modes until $\omega_{\rm F}$ starts to approach $\omega_{\rm D}$, or when $\lambda_{\rm max}$ starts to approach $a$. When $\lambda_{\rm max}=a$, $c_v$ becomes $c_v=2$ according to Eq. (\ref{2}) and undergoes a crossover to another regime given by Eq. (\ref{4}). In this regime, the minimal wavelength of the longitudinal mode supported by the system is $\lambda_{\rm min}=l$. According to Eq. (\ref{4}), $c_v$ remains close to the ideal gas value of $\frac{3}{2}$ in almost entire range of the wavelengths of the longitudinal mode until $\lambda_{\rm min}$ approaches $a$. When $\lambda_{\rm max}=a$, $c_v$ becomes $c_v=2$, and matches its low-temperature value at the crossover as schematically shown in Figure \ref{length}.

Consistent with the above discussion, Figure \ref{length} shows that $c_v$ remains constant at either $3$ or $\frac{3}{2}$ over many orders of magnitude of $\frac{\lambda}{a}$ except when $\frac{\lambda}{a}$ becomes close to 1 by order of magnitude. Figure \ref{length} emphasizes a transparent physical point: modes with the smallest wavelengths comparable to interatomic separations $a$ contribute most to the energy and $c_v$ in the disordered systems (as they do in crystals) because they are most numerous. Consequently, conditions $\lambda_{\rm max}\approx a$ for two transverse modes and $\lambda_{\rm min}\approx a$ for one longitudinal mode, corresponding to the disappearance of modes with wavelengths comparable to $a$, give the largest changes of $c_v$ as is seen in Figure \ref{length}.

The last result is tantamount to the following general assertion not hitherto anticipated: the most important changes in thermodynamics of the disordered system are governed by its {\it fundamental length} $a$ only. Because this length is not affected by disorder, this assertion holds equally in ordered and disordered systems.

We note here that in addition to short-range order, disordered systems also possess medium-range order, the length scale beyond which structural correlations are lost. The absence of order beyond the medium range affects phonon mean free paths and transport properties such as thermal and electric conductivity, yet the changes of thermodynamic properties turn out to be insensitive to structural modifications in the medium range, and are governed by $a$ only.

\subsection{Spin glasses}

Our final example is the calculation of the free energy and susceptibility of disordered magnetic systems, spin glasses, on the basis of collective excitations. Published experimental data related to collective magnetic modes, magnons, are less numerous in spin glasses as compared to phonons in glasses and liquids, presumably due to experimental challenges existing at the time when this area was active, including the need to use polarized neutrons in order to isolate magnetic excitations spread in momentum and energy due to disorder and the associated low signal to noise ratio for the polarized neutrons \cite{lynn}. Nevertheless, approximately linear dispersion relationships were found in spin glass systems experimentally \cite{swe1,swe2,swe3,swe4}, the finding reinforced by theory and modeling \cite{spin}. Using the associated quadratic Debye density of states of magnons and evaluating the free energy and the susceptibility $\chi$ gives \cite{spin}:

\begin{equation}
\chi=\chi_0+\frac{\pi^2}{4}N\mu^2\frac{T^2}{T^3_{\rm D}}
\label{waves}
\end{equation}

\noindent where $T_{\rm D}$ is Debye temperature of spin waves, $\mu$ is magnetic moment and $\chi_0$ is due to zero-point vibrations \cite{spin}. Consistent with experimental $\chi$, this calculation illustrates the operation of collective modes in disordered magnetic systems.

As discussed above, the evidence of propagating collective modes in liquids and glasses in a wide range of $k$ has become available only recently thanks to the deployment of powerful synchrotron radiation sources. We believe it will be interesting to utilize current experimental advances to study spin waves in disordered magnetic systems too, including the increased intensity of neutron beams and increased polarization compared to earlier work \cite{lynn}, particularly in view of the continuing interest in these materials coupled with relative scarcity of experimental data related to spin waves.

\section{Relationship to the previous approach to liquids}

Research into strongly-interacting disordered systems and liquids in particular has a long and rich history spanning hundreds of years. Readers are encouraged to consult general statistical mechanics textbooks as well as textbooks dedicated to liquids \cite{lanstat,ziman,l1,l2,march1,l3,l4,l5,frenkel} for the in-depth discussion of the history of this research and appreciation of the efforts involved. In this review, we have focused on the recent evidence related to collective excitations in disordered systems and liquids in particular, and discussed how this evidence can be used to construct a thermodynamic theory capable of making quantitative predictions of experimental observables such as specific heat shown in Figure 2. Below we comment on how this discussion relates to the previous approach to liquids.

In a crystal, the partition function is evaluated from the Hamiltonian representing normal modes with frequencies extending to the largest, Debye, frequency. Until fairly recently, it has not been evident that high-frequency collective excitations, which make most important contribution to the energy, can propagate in disordered systems and liquids in particular. The long-lived absence of this evidence shaped the historical approach to liquids. In this approach, liquids have been treated as a general many-body statistical mechanics problem where the central task is evaluating the configurational integral \cite{lanstat,ziman,l1,l2,march1,l3,l4,l5,frenkel}. For example, assuming that the interactions and structural correlations are known and pairwise, the liquid energy can be calculated as \cite{ziman,l2,march1,l4,l5,frenkel}:

\begin{equation}
E=\frac{3}{2}NT+\frac{N\rho}{2}\int\phi(r)g(r)4\pi r^2dr
\label{gas}
\end{equation}
\noindent where $\phi(r)$ is interatomic potential, $g(r)$ is radial distribution function and $\rho$ is density.

At this point, we note an interesting difference between Eq. (\ref{1}) and Eq. (\ref{gas}): the former integrates over frequencies of collective excitations whereas the latter integrates over the structure of the disordered system. The traditional approach of evaluating the configurational integral, the second term in Eq. (\ref{gas}), has met widely recognized challenges \cite{l2,march1,frenkel} which we briefly discuss below. These challenges have contributed to the lack of understanding and interpreting the experimental liquid specific heat, resulting in the unfortunate state of affairs summarized by Granato in our introduction.

$\phi(r)$ and $g(r)$ are not generally available for liquids, apart from model systems such as hard and soft spheres, Lennard-Jones systems and similar ones representing simple liquids such as Ar, Kr and so on. For these systems, $\phi(r)$ and $g(r)$ can be determined from experiments or simulations and subsequently used in Eq. (\ref{gas}) to calculate liquid energy and $c_v$, although it is not clear whether this procedure has been shown to reproduce the experimental $c_v$ in Figure 2. Unfortunately, neither $\phi(r)$ nor $g(r)$ are available for liquids with any larger degree of complexity of structure or interactions. Many-body correlations \cite{born,henders} and network effects can be strong in common systems such as olive oil, glycerol, carbon oxide or even water \cite{emilio}, resulting in complicated structural correlation functions that can not be reduced to simple two- or even three-body correlations that are often used in Eq. (\ref{gas}) and similar ones. As widely recognized \cite{l2}, approximations become difficult to control when the order of correlation functions already exceeds three-body correlations. Similarly, it is challenging to extract multiple correlation functions from the experiment. The same problems exist for interatomic interactions which can be equally multi-body, complex and not amenable to determination in experiments or simulations. As a result, evaluating the configurational integral has proved to be a generally intractable task and has not provided insights into the origin of experimental $c_v$ including the behavior shown in Figures 2 and 3.

Contrary to interatomic interactions and structural correlations, $\tau$ is readily available regardless of the liquid complexity: it is either measurable directly or from viscosity measurements, and hence can be used in, e.g., Eq. (\ref{2}), to calculate liquid energy and specific heat. In the approach to liquids based on collective excitations, $\tau$ appears as the relevant physical parameter because it governs the liquid phonon states as proposed by Frenkel. Provided the internal degrees of freedom are excited (as is the case at high enough temperature), the phonon approach to liquid thermodynamics based on $\tau$ predicts the decrease of $c_v$ decreases from 3 to 2 in Figure 2, followed by the different regime where $c_v$ further decreases to its gas value of $\frac{3}{2}$ in Figure 3.

Notably, the traditional approach and our treatment of liquids are fundamentally related and should give the same result for $c_v$. Indeed, $\tau$ ($\omega_{\rm F}$) in Eq. (\ref{2}) is governed by structure ($g$) and interactions ($\phi$) in Eq. (\ref{gas}). A recent computer simulation of simple model systems has shown that liquid structural, thermodynamic and dynamic properties are all inter-related, including at the Frenkel line crossover \cite{jchem}. However, the dependence of $\tau$ on $g$ and $\phi$ is complex (noting the caveat that $g$ and $\phi$ are intricately and often non-trivially inter-related themselves), because $\tau$ is defined by the activation barriers and not by the equilibrium properties.

Generally, the activation barriers and $\tau$ can be evaluated in the computer simulation for given $g$ and $\phi$ in a fairly small system of the size on the order of hundred atoms, but the evaluation is challenging for realistic systems because interactions and structural correlations are not known except for a small set of simple systems. Analytically, the calculation of activation barriers and $\tau$ becomes intractable even for a small number of particles due to the complexity of treating the effects of non-linearity. Indeed, the system of two coupled anharmonic oscillators with simple model forms of anharmonicity can be solved exactly, with the result that hopping between different equilibrium positions emerges as a result of a set of bifurcations, with barrier heights and $\tau$ directly related to the non-linearity parameters in the interaction potential \cite{nonlinear}. This constitutes an important first-principles proof that hopping emerges as a result of bifurcations in the non-linear system. Unfortunately, increasing the number of agents above 2--3 or altering the simple analytical form of non-linearity makes the analysis impossible even using approximate schemes \cite{nonlinear}.

It is here where the introduction of $\tau$ provides an enormous reduction of the complexity of the problem discussed above. We call it ``Frenkel reduction''. In this reduction, $\tau$ encompasses the physically relevant and important features of the unknown complex energy landscape in one simple and available physical parameter. Governing the number and type of propagating collective excitations in the liquid, $\tau$ serves as the main property of the liquid relevant for most of its thermodynamic properties.

An interesting objection could be raised that although our approach explains the experimental $c_v$ of liquids in Figure 2, the approach is based on $\tau$, the emergent property, but not on more fundamental condensed matter properties such as $g(r)$ and $\phi(r)$, and hence is not a ``first-principles'' description of liquid thermodynamics and $c_v$. This brings us to an important question of what we aim to achieve by a physical theory. According to one view, ``The point of any physical theory is to make statements about the outcomes of future experiments on the basis of results from the previous experiment'' \cite{landau-pi}. This emphasizes relationships between experimental properties. In this sense, Eqs. (\ref{2}) and (\ref{4}) provide relationships between liquid thermodynamic properties such as energy and $c_v$ on one side and its dynamic and oscillatory properties such as $\tau$ or $l$ on the other.

\section{Summary and outlook}

The first main point of this discussion is that disordered systems support eigenstate collective excitations that are non-decayed and non-damped as in ordered crystals, the point which is important to emphasize in view of previous beliefs. We have noted that disorder can have notable consequences for system's properties, including transport. On the other hand, thermodynamic properties such as energy, heat capacity, susceptibility and so on are not sensitive to disorder; namely, these properties turn out to be insensitive to the difference between the eigenstate collective modes of the disordered system and harmonic plane waves. We have illustrated how this assertion provides a way in to calculate thermodynamic properties of different disordered systems and liquids in particular, opening a new line of enquiry into this long-standing problem. An interesting insight from the recent work is the realization that most important changes of thermodynamic properties of the disordered system
are governed only by its fundamental length, the interatomic separation.

K. T. is grateful to EPSRC, and V. V. Brazhkin to RFBR for financial support. The authors are grateful to Yu. D. Fomin and A. G. Lyapin for discussions.

\end{document}